# Socially-Aware Venue Recommendation for Conference Participants

Feng Xia[1], Nana Yaw Asabere[1], Joel J.P.C Rodrigues[2], Filippo Basso[2], Nakema Deonauth[1], Wei Wang[1]
[1]School of Software, Dalian University of Technology, Dalian 116620, China
[2]Instituto de Telecomunicações, University of Beira Interior, Portugal
*f.xia@ieee.org; yawasabere2005@yahoo.com; joeljr@ieee.org; f.basso@it.ubi.pt; aether46@gmail.com; ehome.wang@outlook.com*

*Abstract*—Current research environments are witnessing high enormities of presentations occurring in different sessions at academic conferences. This situation makes it difficult for researchers (especially juniors) to attend the right presentation session(s) for effective collaboration. In this paper, we propose an innovative venue recommendation algorithm to enhance smart conference participation. Our proposed algorithm, Social Aware Recommendation of Venues and Environments (SARVE), computes the Pearson Correlation and social characteristic information of conference participants. SARVE further incorporates the current context of both the smart conference community and participants in order to model a recommendation process using distributed community detection. Through the integration of the above computations and techniques, we are able to recommend presentation sessions of active participant presenters that may be of high interest to a particular participant. We evaluate SARVE using a real world dataset. Our experimental results demonstrate that SARVE outperforms other state-of-the-art methods.

## I. INTRODUCTION

Informally an event is an organized situation period that is only valid for a short period of time. Events such as smart conferences, smart meetings, symposia and workshops are regularly organized worldwide each year. The organizing process consists of classifications of major activities involving several distant participants. Due to the characteristics of these events, distributed solutions are suitable for their management.

Academic conferences and workshops do not just serve as platforms to present the research work of participants, but also aim to connect researchers/participants in the same domain and foster prospective collaborations. Different participants or attendees at these events are likely to have diverse research interests within extensive research disciplines [1].

The plethora of talks and presentations in multiple and parallel tracks at academic conferences makes it difficult, especially for junior researchers to attend the right presentation sessions and collaborate socially with participants and potential researchers who have similar research interests. Additionally, because smart academic conferences are vibrant, participants are likely to find themselves moving around and attending different talks in different rooms and at different times, as a result of the uncertainty of which particular and reliable presentation session(s) to participate or attend. The main program schedule of conferences can also change, for instance the scheduled presentation sessions may be canceled due to the non-attendance of the presenter.

Recommender systems are software applications that attempt to reduce information overload by recommending items of interest to end users based on possible preferences such as movies, books and other relevant products/places [2]. Therefore, depending on the scope of events such as smart conferences, a mobile multimedia recommender system incorporated with context and social awareness is necessary to generate effective and reliable presentation sessions at the conference for attendees/participants.

In this paper, we corroborate the importance of social event participation by improving the activities of conference participants in relation to the location of communities involving presentation session venues. We are encouraged and motivated to believe that participation in smart conferences can be enhanced through the integration of mobile technological devices, recommender system techniques, contextual information and social properties which will effectively enhance social awareness at such events. Such a novel approach will efficiently enable the achievement of high social capital and successful social learning at smart conferences.

Furthermore, we propose an innovative solution called Social Aware Recommendation of Venues and Environments (*SARVE*). Through mobile device technology, *SARVE* recommends conference presentation session venues and environments to participants by utilizing socially aware, and distributed community detection techniques. The main aim of *SARVE* is to detect and recommend conference presentation session venues that are important and related to the research interests of participants. This will enable effective participation in conference session talks and presentations facilitated by other researchers (presenters) of similar interests and high popularity.

Initially, our proposed *SARVE* explicitly obtains information concerning the research interests, contact durations and contact frequencies of individual conference participants in order to determine their preference similarities and social tie strengths in terms of research. To detect different communities consisting of presentation sessions at the conference, *SARVE* computes and employs different sources of information. These include: (i) context (locations and times of different presentation sessions and available times and locations of participants), (ii) personal (research interests of participants) and (iii) social (tie strength between the presenters and the other conference participants and degree centrality of the presenter). The distributed community detection algorithm which is utilized in *SARVE* is responsible for organizing the conference participants into different and common communities/sessions pertaining to talks and presentations at the conference.

The rest of the paper is organized as follows. Section II reviews related work on social and contextual recommendations. The architecture and algorithmic design of our SARVE are discussed in Sections III and IV respectively. In Section V, we present our experimental evaluations. Finally, Section VI concludes the paper.

## II. RELATED WORK

Quite a number of recommender systems and algorithms involving the utilization of contextual and social information have been presented and discussed by various researchers in recent years. A significant number of these solutions were proposed to recommend items/venues/places based on the combination of both user/item interest and contextual information [3] or user/item interest and social information [4]. Others such as [1][3]-[7] were also proposed to recommend items/venues/places based on a combination of context, user/item interest and social information retrieved through techniques such as matrix factorization, social networking analysis and data mining.

Baltrunas *et al.* [3] took a new approach for assessing and modeling the relationship between contextual factors and item ratings. Instead of using the traditional approach of data collection, where recommendations are rated with respect to real situations in which participants go about their normal lives, they simulated contextual situations to capture data more easily in terms of how context influences user ratings.

Mohsen *et al.* [4] introduced a generalized stochastic block model called *GSBM*. *GSBM* models not only the social relations but also the rating behavior. *GSBM* also learns the mixed group membership assignments for both users and items in a Social Rating Network (SRN).

Biancalana *et al.* [5] described a social recommender system that is able to identify user preferences and information needs, thus suggesting personalized recommendations related to Point of Interests (POIs) in the surroundings of the user's current location context. Similar to [5], Beach *et al.* [6] presented a system called *Whozthat*, which is also based on user's current location. *Whozthat* ties together online social networks with mobile smartphones to answer the common and essential social question (who is that). *Whozthat* offers an entire environment in which increasingly complex context-aware applications for recommendations can be built.

In terms of recommendations involving presentation/talk session venues at conferences, Pham *et al.* [1] presented the Context-Aware Mobile Recommendation Services (*CAMRS*), which is based on the current contexts (whereabouts at the venue, popularity and activities of talks and presentations) sensed at the conference venue. Similar to [1], Farzan and Brusilovsky [7] presented a social information access system that helps researchers attending a large academic conference to plan talks they wish to attend.

Our *SARVE* approach seeks to utilize not only the Pearson correlations obtained through tag rating preference similarities of conference participants, but also their social ties computed through contact durations and contact frequencies. Furthermore, we integrate contextual information consisting of different locations and times pertaining to the conference venue and participants as well as the degree centrality (popularity) of presenters.

To the best of our knowledge, the generation of social recommendations for conference participants using a combination of Pearson correlation, social ties, contextual information and the degree centrality (popularity level) of a presenter is quite uncommon. None of the above described categories of related work and methods incorporate social properties in their recommendation processes/procedures. Additionally, we are inspired to embark on this research because the social properties of nodes/users in a network are important factors to consider when analyzing social data for an effective output such as recommendation.

## III. SARVE FRAMEWORK

As previously mentioned, the tremendous number of presentation session venues at smart conferences and the diverse interests of participants/attendees, has resulted in the difficulty of locating precise presentation session venues. This has become a significant problem for conference participants (especially young or new researchers).

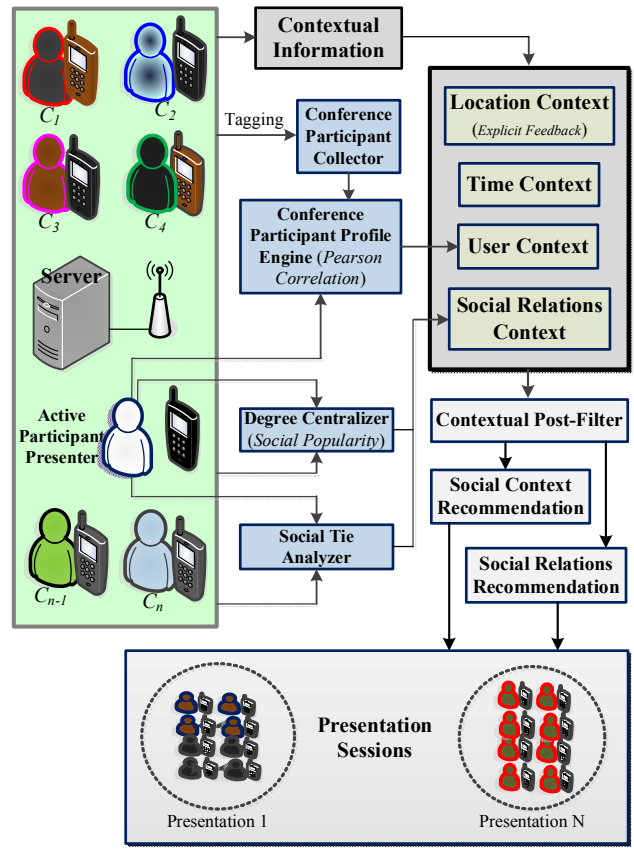

Figure 1. SARVE Framework

This section presents the basic idea and framework of *SARVE*. Figure 1 shows that, by the augmentation of relevant context, our *SARVE* framework generates social relations recommendation through social tie computations of participants (i.e. social ties of $C_p$ and other participants) and degree centrality of the presenter. *SARVE* generates social context recommendation through the Pearson correlations of the participants (i.e. Pearson correlations between $C_p$ and other participants).

The left hand side of Figure 1 depicts an interactive scenario of the conference participants ($C_1$.......$C_n$), who are the users and an active participant presenter ($C_p$) at the smart conference. Referring to Figure 1, if a conference participant makes a social recommendation request to attend a relevant presentation session(s) at a particular conference venue, *SARVE* computes the Pearson correlation and social ties of the user and all the presenters to ascertain high levels of similarity and ties strength between them. Additionally, *SARVE* further computes degree centrality of participant presenters to determine their popularity status/level at the smart conference and further integrates explicit contextual information of the user, presenters and community, in order to accordingly generate an effective social venue recommendation.

The various components of our *SARVE* framework are described below. In Figure 1, the *Conference Participant Collector* gathers and sends the collaborative tag ratings of the individual conference participants to the *Conference Profile Engine* for the computation of user context.

The *Social Tie Analyzer* computes the contact durations and contact frequencies between $C_p$ and the other conference participants to determine their tie strengths. For example, if a conference participant (active user) specifies that his contact frequency at the conference with a $C_p$ is 6 in a duration of 70 minutes and conference time frame of 720 minutes, then using (3), the social ties result will be computed as 0.58. The *Degree Centralizer* computes the social popularity of a $C_p$ with other conference participants by measuring the extent of their direct social links and ties.

The *Contextual Post-Filtering* technique involves contextualizing recommendation outputs for each conference participant based on their tagged ratings through traditional 2D procedures of the entire data [8]. Therefore, through the *Contextual Post-Filter*, *SARVE* verifies and contextualizes the resultant location, time, user and social relations contexts of the smart conference community and participants. We elaborate further on the algorithmic design of our *SARVE* framework in next section.

## IV. ALGORITHMIC DESIGN OF SARVE

In this Section, we present a description of our algorithmic design for the *SARVE* framework shown in Figure 1. We firstly describe our approach of computing similar research interests of conference participants and participant presenters using Pearson correlation. Then, in the next two subsections, we describe our methods for computing the social ties of the participants and degree centrality of the participant presenters. In the last subsections, we describe how we sense contextual information and match contextual relationships in *SARVE*.

### A. User Interest and k Most Similarity

We propose an explicit approach involving conference participants specifying their research interests by using their mobile devices to input specific keywords in the form of tags to denote interest in some specific topics/research disciplines. Furthermore, the contact durations and frequencies between the presenters and other conference participants are also obtained explicitly from the individual participants (users). In our proposed algorithm, a tag is a relevant keyword assigned to one or more research interests of a conference participant, which describes and gives more ideas about a research area and enables it to be classified.

Traditionally, recommender algorithms can be categorized into three main traditional categories, these include [9]: Collaborative Filtering (CF), Content-Based Recommendation (CBR) and Hybrid Recommendation (HR). CF is the most successful recommender system and can give more accurate recommendations in comparison to *CBR*, since it is more beneficial in terms of a user's personal tastes/interests.

Because *SARVE* consists of a user-item database, we utilize the memory-based CF approach with a focus on user-based CF (user similarity in terms of research interests), which involves the following steps:
1. Look for users (participant presenters) who share the same rating patterns with the active user (conference participant - the user whom the recommendation is for).
2. Use the ratings from those like-minded and similar interest users found in Step 1 to calculate a recommendation for the active conference participant.

To perform the above steps, we utilize Pearson correlation to identify and compute the *k* most similarity between two users' (nearest neighbors) involving a participant presenter, $C_p$ and a conference participant, $C_x$. Each user is treated as a vector in the *m*-dimensional item space and the similarities between $C_p$ and $C_x$ are computed within the vectors.

$$Sim(c,d) = \frac{\sum_{i \in I}(r_{c,i} - \bar{r}_c)(r_{d,i} - \bar{r}_d)}{\sqrt{\sum_{i \in I}(r_{c,i} - \bar{r}_c)^2} \sqrt{\sum_{i \in I}(r_{d,i} - \bar{r}_d)^2}} \quad (1)$$

Pearson correlation measures the extent to which two variables linearly relate with each other. After the *k* most similar users have been identified through the user-item matrix, user-based CF techniques then generates a top-*N* recommendation list for $C_x$ based on similarities with $C_p$. For the user-based CF algorithm, the Pearson correlation between conference participants, $C_p$ and $C_x$ is computed using (1).

In (1), $C_p$ and $C_x$ are represented as *c* and *d* respectively. Therefore the similarity between $C_p$ and $C_x$ is denoted by *Sim(c, d)*. The tagged ratings of *c* and *d* for item *i*, (whereby *i* ∈ *I* and *I* is the set of items) are denoted by $r_{c,i}$ and $r_{d,i}$ respectively. The average ratings of *c* and *d* are denoted by $\bar{r}_c$ and $\bar{r}_d$ respectively. Using (2), we set a threshold, *γ* (to be determined in our experiment) for (1), to define the preference similarity between $C_p$ and $C_x$ in terms of tagged (keyword) ratings (scores 1-5).

$$Sim(C_p, C_x) \geq \gamma \qquad (2)$$

The similarity values between $C_p$ and the other conference participants has to fall within the defined threshold before such participants can be detected as members of the community where the participant presenter will be delivering his/her presentation.

### B. Tie Strength

The social relations between individuals are usually called social ties. Ties typically represent the existence or non-existence of a substantial relationship between two individuals, for example, acquaintance, research familiarities etc [10]. We measure and estimate the tie strength between a $C_p$ and $C_x$ using (3).

$$SocTie_{C_p, C_x}(t) = (\lambda_{C_p, C_x} \times d_{C_p, C_x}(t))/T \qquad (3)$$

In (3) $d_{C_p, C_x}(t)$ is the contact duration between the participant presenter, $C_p$ and another conference participant, $C_x$ in the time frame $T$ and $\lambda_{C_p, C_x}$ is their contact frequency (i.e. the number of times $C_p$ and $C_x$ have been in contact within the time frame $T$).

To define the tie strength between $C_p$ and $C_x$, we set a threshold, $\beta$ (to be determined in our experiment) for (3) using (4). The social tie values between $C_p$ and the other conference participants has to fall within the defined threshold before such participants can be detected as members of the community where the participant presenter will be delivering his/her presentation.

$$SocTie_{C_p, C_x}(t) \geq \beta \qquad (4)$$

### C. Degree Centrality

Degree Centrality measures the numbers of direct ties that is involved with a given user/node. Consequently, a user involved with more social ties represents a more important location for a community in a network than a user with fewer or no social ties. A user with high degree centrality maintains contact durations and frequencies with other users within the network. Such users can be seen as the most active and popular with a large number of links to other users in the same network [10][11].

Therefore, in *SARVE*, we assume that participant presenters ($C_{Ps}$) that have a higher number of social ties with other participants ($C_x$) are popular and consequently their popularity can be used as added incentives to generate effective presentation session recommendations for the conference participants. Moreover, $C_{Ps}$ that maintain few or no social ties and relations are described as unpopular within the network. The degree centrality for a given $C_P$, includes a function $a$, where $a(C_P, C_x) = 1$, if a direct link exists between $C_P$ and $C_x$. Degree centrality for a given $C_P$ is therefore computed as:

$$C_D(C_p) = \sum_{k=1}^{N} a(C_p, C_x) \qquad (5)$$

where $N$ is the total number of users/nodes in the network.

### D. Contextual Information Sensing

A specific definition and model of context in recommender systems can expedite what does and does not constitute context and can facilitate the usage of contextual data across various applications. From an operational perspective, context is often defined as an aggregate of various categories that describe the setting in which recommender systems are deployed. Some examples of applicable contexts in recommender systems include: location, current activity, available time of users, physical conditions and social relations [8][12]. To this extent, *SARVE* utilizes four types of contexts namely, location, time, user and social relations. We describe how these contexts are sensed by *SARVE* below.

*Location Context*: Location context has dominated research in context-aware recommender systems and mobile computing on a large scale [8][12]. Location models and sensors such as Global Positioning System (GPS) and Wi-Fi that capture geometric information of objects that are human-readable have been proposed by various researchers. Location context in recommender systems is often acquired or sensed implicitly through GPS or Wi-Fi location sensors in mobile devices [13]. Other recommender systems which require the exact location information usually rely on explicit approaches such as [14]. With reference to the above described scenarios, *SARVE* involves the detection of exact venues of presentation sessions. Therefore, we utilize an explicit procedure to sense the precise locations of participant presenters and the other participants at the smart conference.

*Time Context*: Time context usually involves the exact date and time information for recommender system scenarios. Time can either be specific (e.g. within five minutes) or imprecise (e.g. within a week, sometime in a month or in the coming semester/academic year). Time and other contexts such as location are usually combined to provide effective recommendations using efficient and innovative tools such as timestamp or time span [12]. In most cases, timestamp data is captured implicitly from available data such as a learning schedule. For example, in [15], the Context-aware Adaptive Learning Schedule (CALS) provides a learning schedule that allows users to enter their time data, in order for them to make plans for their leaning activities. Similar to CALS, *SARVE* also provides a smart conference schedule with available presentation session dates and times to enable users enter their specific time data for available presentation sessions.

*User Context*: As described above, we sense the context of the users (participant presenters and conference participants) through explicit collaborative tagging of their research interests.

*Social Relations Context*: As presented above, we sense the social relations context of the participant presenters and the other conference participants through the computation of their social ties and degree centrality.

### E. Contextual Relationship Matching

In social tagging systems, a user's tagging and commenting activities generate relations involving more than two types of entities [16] and the posts (that is, each tag produced by a user for an item) are classified as third order data [17]. Yin *et al.* [16] highlighted that this classification is further considered as a triple (user-tag-item) as shown in Figure 2. We adopt the model called the *Bipartite graph between relations and entity types* in [16] and use it to create social relationships between $C_p$ and $C_x$ in terms of context. This will enable the generation of effective and efficient social recommendations based on the *k* most similarity and social tie results of participants obtained from (1) and (3) and subsequent computed threshold values from (2) and (4). An example of four relations on five entity types in a social tagging system is depicted in Figure 2.

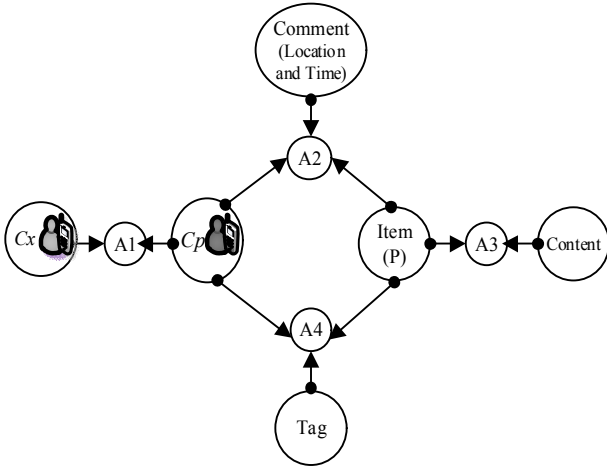

Figure 2. Bipartite graph between conference participant relations and entity types

In Figure 2, A1 is the social network context (user-user), A2 is the comment context (user-comment-item), A3 is the item-content context (item-content feature) and A4 is the tag post context (user-tag-item). If the results of (1) and (3) depict that $C_p$ and $C_x$ have *k* most similarities and strong social ties, then the presentation (*Item (P)*) annotated with a tag by $C_p$, based on a comment feature about the location and time of the presentation and content feature will be the recognized and detected presentation community for $C_x$. It must be noted that the extent of social relationship in terms of context between $C_p$ and $C_x$ can only be generated based on the results of (1) and (3) i.e. if the research interest similarities and social ties of $C_p$ and $C_x$ doesn't fall within the computed threshold results, a social relationship cannot be established using Figure 2.

*F. Community Detection*

The major problem for a community structure to be introduced in a Mobile Social Network (MSN), lies in the community detection algorithm. There are two types of methods used to detect a community in MSNs, these are centralized and distributed community detection techniques. In the *centralized* technique, full knowledge of the whole MSN and its ties are needed, while in the *distributed* technique, each node or user is able to detect the community it belongs to [10].

We propose a novel distributed community detection algorithm in which users (conference participants) independently detect related presentation session venues (communities) through the generation of social context recommendations and social relation recommendations pertaining to the presentation session venues of participant presenters. We detect distributed communities of conference participants based on their research interests, social ties and social/collaborative tag ratings as well as the social popularity of participant presenters.

Our proposed distributed community detection algorithm initially declares and initializes integer, floating and string variables. The integer variables consist of *i, j, m, n* and *z*, whereby *i* and *j* are initialized to a value of 0 and used for comparison of transactions in the arrays of participant presenters [m] and conference participants [n] through *for* loops based on incremental transactions. These steps are depicted in 1-8. Steps 9 and 10 compute the Pearson correlations between the conference participants and presenters and compares the results to a threshold value. Based on the results of the Pearson Correlation computation, steps 11-17, compare the contextual parameters of conference participants and presenters and accordingly generates social context recommendations. The final steps (18-28) of our proposed algorithm, compute both the social

---

**Algorithm**: Pseudocode for detecting and recommending presentation session venues

1: // Declare and Initialize Variables
2: *i, j, m, n*, and *z*;             // integer variables
3: *pearson_threshold_val, soctie_ threshold_val, social_tie[z] deg_cent_threshold* and *Pearson[z]*;     // floating variables
4: *location[n], time[n]*;     // string variables
5: Participants [n];          // array of Participants of size n
6: Presenters[m];            // array of Presenters of size m
7: **for** (i=0 to i<n increment i)
8:  **for** (j=0 to j<m increment j)
9:   Compute Pearson correlations using Eq. (1) and store in Pearson[z]
10:   **if** (Pearson[z] ≥ pearson_threshold_val) **then**
11:    Compare contextual parameters;
12:    **if** (Presenter[j].location = Participant[i].location) **AND** (Presenter[j].time = Participant[i].time) **then**
13:     // Generate Social Context Recommendation
14:     Assign Participant[i] to Presenter[j];
15:    **end if**
16:   **end if**
17:   increment z
18:   Compute Social Ties using Eq. (3) and store in social_tie[z]
19:   Compute Degree Centrality of Presenters using Eq. (5)
20:   **if** (*SocTie$_{Cp, Cx}(t)$* ≥ soctie_ threshold_val) **OR** (Participant[j].deg_cent≥deg_cent_threshold) **then**
21:    Compare contextual parameters;
22:    **if** (Presenter[j].location = Participant[i].location) **AND** (Presenter[j].time = Participant[i].time) **then**
23:     // Generate Social Relations Recommendation
24:     Assign Participant[i] to Presenter[j];
25:    **end if**
26:   **end if**
27:  **end for**
28: **end for**

ties of the conference participants and presenters and the degree centrality of presenters and accordingly generates social relation recommendations.

## V. EXPERIMENTAL EVALUATION

In order to validate our *SARVE* approach described in the previous sections, this section presents the performance and evaluation of relevant benchmarking experiments. The subsections below describe our experimental procedure.

### A. Dataset and Experimenatl Setup

The two main types of experimental procedures used for evaluating recommendation algorithms are referred to as online and offline evaluations. Many real world systems utilize an online testing system where multiple algorithms can be compared with each other to test their performance. However, online evaluation experiments in a multitude of cases are very expensive/costly and also stand the chance of test users being initially discouraged from using the system in real time since there is no opportunity to evaluate algorithms before presenting results to users. We therefore adopted an offline evaluation procedure for our experiment [18].

Our goal was therefore to simulate an online process where the system makes recommendations or predictions and the user uses the recommendations or corrects the predictions. We therefore simulated the 2012 International Conference on Web-Based Learning (ICWL 2012) which involved recording historical user data in order to obtain the knowledge of how a user will rate an item or which recommendations a user will act upon.

To identify research interests and contextual information of conference participants which will be compared to the interests and contextual information of the presenters, we gathered data from 78 members/students of the School of Software, Dalian University of Technology, China. Explanation was given to members/students to select/annotate keywords of interest as well as contextual information (available time and present location) in relation to the simulated conference (ICWL 2012). We divided the above described dataset into two parts namely, the training set (80%) and the test set (20%). Figure 3 shows more information about the data simulated from ICWL 2012.

The highest contact durations and contact frequencies for the ICWL 2012 simulation dataset were 80 minutes and 7 respectively i.e. $dc_{p,c_x}(t) = 80$ and $\lambda c_{p,c_x} = 7$. Empirically, we assumed a time frame *T* of 12 hours (720 minutes) for the total duration of the smart conference. Using (4), we computed $SocTiec_{p,c_x}(t) = (80 \times 7)/720$ and obtained a result of 0.8 as the highest positive and effective recommendation based on strong social ties between participant presenters and other participants. Consequently, we set the range for recommendation based on the social ties (relations) as $0 \leq SocTiec_{p,c_x}(t) \leq 0.8$ and allocated a social relations recommendation threshold of 0.5 and above in accordance to the dataset.

Schafer *et al.* [19] emphasized that Pearson correlation ranges from 1.0 for users with perfect agreement to -1.0 for perfect disagreement users i.e. $-1 \leq$ *Pearson correlation* $\leq 1.0$. Negative correlations are generally believed to be invaluable in increasing prediction accuracy and recommendation and hence should be ignored for effective recommendations. We therefore observe in our experiment that, a Pearson correlation value of 1 between presenters and participants signifies high *k* most similarities of research interests for onward comparison of contextual parameters and the effective generation of social context recommendations.

### B. Evaluation Metrics

In recommender systems research, it is extensively assumed that a recommendation is successful if and only if the recommended item/resource is beneficial and also if and only if the item preference matches the target user's preferences. To this extent, we adopted two commonly used classification metrics, Precision and Recall.

$$Precision = \frac{e}{e+f} = \frac{good\ venues\ recommended}{all\ venue\ recommendations} \quad (6)$$

$$Recall = \frac{e}{e+g} = \frac{good\ venues\ recommended}{all\ good\ venues} \quad (7)$$

Precision metrics measures a recommender algorithm's ability to show only useful items, while it tries to minimize a combination of them with useless ones. Recall metrics measures the coverage of useful items/resources the recommender algorithm/system can achieve. In other words, recall metrics measures the capacity of a recommender system/algorithm to obtain all useful items/resources present in the pool [18]. Olmo and Gaudioso [18] summarized these facts using the confusion matrix in Table I. Equations (6) and

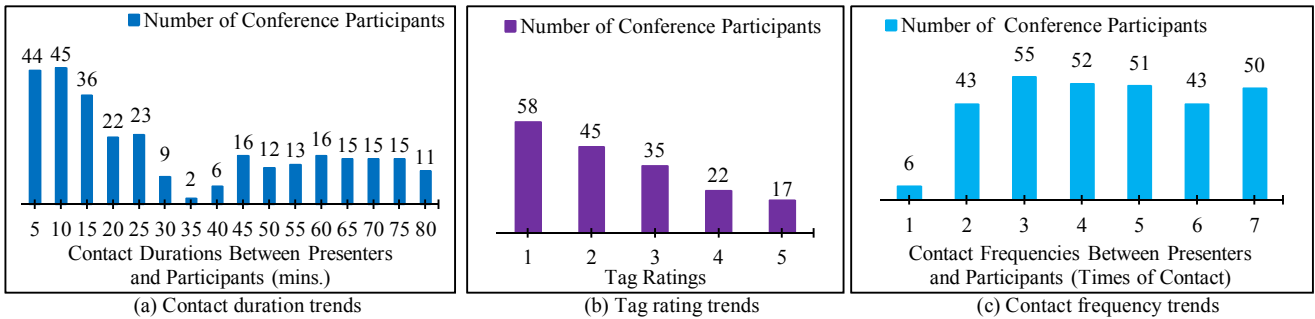

Figure 3. Details and components of ICWL 2012 dataset

(7) respectively depict the computations of precision and recall using variables *e, f* and *g* in Table I.

### C. Experimental Results and Analysis

In order to authenticate the performance of our proposed algorithm (*SARVE*) using the above evaluation metrics, we compared *SARVE* to the work done by Pham *et al.* [1] and Farzan and Brusilovsky [7]. In our experiment, B1 and B2 denote the methods of [1] and [10] respectively. As elaborated in Section II, both B1 and B2 involved recommendations for conferences presentation sessions which are quiet similar and related to *SARVE*.

In terms of precision, both social context and social relations recommendations for *SARVE* were more precise and exact especially at higher recommendation values in accordance to the dataset. Referring to Figure 4(a), at the highest value for Pearson correlation (1.0), *SARVE* attained a higher precision (0.096) which was higher in comparison to that of B1 (0.075) and B2 (0.045). Similarly, in Figure 4(b), at the highest value for social ties (0.8), *SARVE* achieved a higher precision of 0.013 in comparison to that of B1 (0.0013) and B2 (0.0011). These scenarios in our experiment substantiates the fact that *SARVE* showed the capacity to show/display more useful and exact items (presentation session venues) in comparison to B1 and B2.

In terms of recall, both social context recommendation and social relations recommendation for *SARVE* exhibited higher recall values and covered more useful items in accordance to the dataset. Referring to Figure 5(a), at the highest value for Pearson correlation (1.0), *SARVE* attained a recall value of 0.810 in comparison to B1 (0.759) and B2 (0.698). Correspondingly, in Figure 5(b), at the highest value for social ties (0.8), *SARVE* attained a higher recall (0.809) in comparison to that of B1 (0.769) and B2 (0.728). This verifies that in our experiment, *SARVE* was able to execute a higher coverage of useful items (presentation session venues) within the pool in comparison to B1 and B2.

The *SARVE* proposed in this paper utilizes socially-aware recommendation through the integration of some social properties of the conference participants. In comparison to B1 and B2, *SARVE* establishes a community detection approach for presentation session venues at the smart conference. Due to the effective utilization of contextual and social characteristic information pertaining to the smart conference environment, it can vividly be seen that our algorithm outperforms both B1 and B2. Both B1 and B2 utilize Pearson correlation and B1 also utilizes social network analysis and link prediction, but the incorporation of the social properties in our experiment validates our innovative approach. Further evidence of our experimental results are statistically depicted in Tables II and III.

Additionally, because our approach generates both social relation and social context recommendations, it is not totally reliant on tag ratings of participants, thus reducing a significant amount of data sparsity. Therefore, by supporting user ratings and ensuring that participants are connected through a network of trust and social relationships our method reduced cold start problems (new user and new item problems).

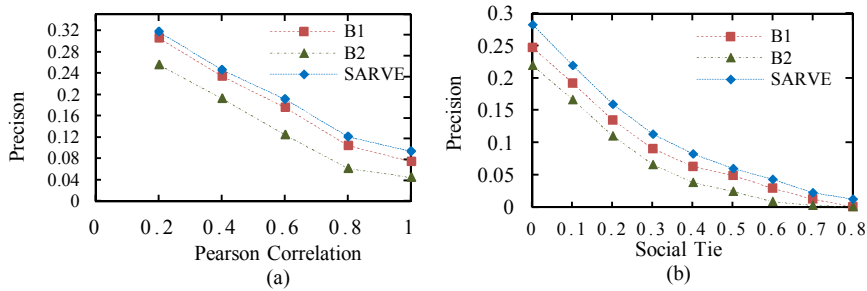

Figure 4. Precision performance for ICWL 2012 dataset (a) social context recommendation (b) social relations recommendation

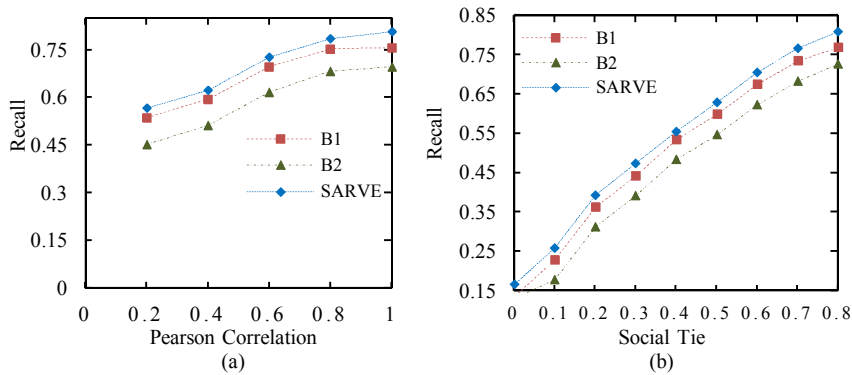

Figure 5. Recall performance for ICWL 2012 dataset (a) social context recommendation (b) social relations recommendation

TABLE I
CONFUSION MATRIX OF TWO CLASSES WHEN CONSIDERING THE RETRIEVAL OF ITEMS

| Classes | Relevant | Not Relevant |
|---|---|---|
| Retrieved | *e* | *f* |
| Not Retrieved | *g* | *h* |

TABLE II
COMPARISON OF PROPOSED ALGORITHM IN TERMS OF PRECISION AND RECALL FOR SOCIAL CONTEXT RECOMMENDATION

| Algorithm | Highest Pearson | Precision | Recall |
|---|---|---|---|
| B1 | 1.0 | 0.075 | 0.759 |
| ***SARVE*** | ***1.0*** | ***0.096*** | ***0.810*** |
| B2 | 1.0 | 0.045 | 0.698 |

TABLE III
COMPARISON OF PROPOSED ALGORITHM IN TERMS OF PRECISION AND RECALL FOR SOCIAL RELATIONS RECOMMENDATION

| Algorithm | Highest Social Tie | Precision | Recall |
|---|---|---|---|
| B1 | 0.8 | 0.0013 | 0.769 |
| ***SARVE*** | ***0.8*** | ***0.013*** | ***0.809*** |
| B2 | 0.8 | 0.0011 | 0.728 |

## VI. Conclusion

This paper has presented a socially-aware recommendation approach that can be used to improve smart conference participation. We proposed a novel solution called *SARVE*, which recommends presentation session venues for participants at a smart conference. Using data consisting of context, social characteristics and research interests obtained through a relevant dataset, we were able to identify neighbors (participants who have similar interests and targets). We used this information as a guide to detect relevant communities pertaining to presentation session venues at the smart conference for the users (participants).

The results of our experiment depict that our approach is capable of providing useful social recommendations to conference participants and outperforms other state-of-the-art methods. In the future, we would like to use other evaluation metrics to assess *SARVE* in more smart conferences. This will enable the verification of different impacts involving recommender information on the quality of social recommendations. To achieve this target, location and proximity sensing instruments as well as computation of other social properties must be researched and explored.

## Acknowledgment

This work was partially supported by the Natural Science Foundation of China under Grant No. 60903153 and No. 61203165, Liaoning Provincial Natural Science Foundation of China under Grant No. 201202032, the Fundamental Research Funds for the Central Universities, the Instituto de Telecomunicações, Next Generation Networks and Applications Group (NetGNA), Portugal, and National Funding from the FCT – Fundação para a Ciência e Tecnologia through the Pest-OE/EEI/LA0008/2013 project.